\documentstyle[12pt,epsfig]{article}
\advance\topmargin by -5\baselineskip
\advance\textheight by 4\baselineskip
\begin{document}
\newcommand{\be}{\begin{equation}}
\newcommand{\ee}{\end{equation}}
\newcommand{\bea}{\begin{eqnarray}}
\newcommand{\eea}{\end{eqnarray}}
\newcommand{\btab}{\begin{tabular}}
\newcommand{\etab}{\end{tabular}}
\newcommand{\bef}{\begin{figure}}
\newcommand{\eef}{\end{figure}}
\newcommand{\bt}{\begin{table}}
\newcommand{\et}{\end{table}}
\newcommand{\ben}{\begin{enumerate}}
\newcommand{\een}{\end{enumerate}}
\newcommand{\ba}{\begin{array}}
\newcommand{\ea}{\end{array}}
\newcommand{\twowaypartial}{i \!\!\stackrel{\leftrightarrow}{\partial}}
\newcommand{\half}{\frac{1}{2}}
\newcommand{\simle}{\stackrel{<}{\sim}}
\newcommand{\simleq}{\stackrel{<}{\sim}}
\newcommand{\simgeq}{\stackrel{>}{\sim}}
\newcommand{\slashp}{\not \!p}
\newcommand{\slashq}{\not \!q}
\newcommand{\slashk}{\not \!k}
\newcommand{\slashepsilon}{\not \!\epsilon}
\newcommand{\slashP}{\not \!\!\!\:P}
\newcommand{\slashQ}{\not \!\!\!\:Q}
\newcommand{\slashK}{\not \!\!\!\:K}
\thispagestyle{empty}
\begin{titlepage}
\begin{center}
\vspace{0.5in}
{\Huge Relativistic and Binding Energy Corrections
To Direct Photon Production In Upsilon Decay}\\
\vspace{1.0in}
{\bf Mohammad Ali Yusuf}   \\
Department of Physics    \\
Quaid-e-Azam University  \\
Islamabad, Pakistan. \\
and    \\
{\bf Pervez Hoodbhoy}   \\
Center for Theoretical Physics  \\
Massachussets Institute of Technology   \\
Cambridge, MA 02139, USA   \\
and  \\
Department of Physics    \\
Quaid-e-Azam University  \\
Islamabad, Pakistan.
\end{center}
\vspace{0.5in}
\begin{abstract} 
A systematic gauge-invariant method, which starts directly from QCD, is 
used to calculate the rate for an upsilon meson to decay inclusively
into a prompt photon. An expansion is made in the quark relative velocity 
$v$, which is a small natural parameter for heavy quark systems. Inclusion 
of these 
O$(v^2)$ corrections tends to increase the photon rate in the middle z 
range and to lower it for larger z, a feature supported by the data.
\end{abstract}
\end{titlepage}
\newpage
\section*{INTRODUCTION}
The hadronic decays of the $\Upsilon$ family of $b \bar{b}$ mesons proceeds
mainly via an intermediate state consisting of three gluons. By replacing one 
of the outgoing gluons with a photon one obtains the leading order 
contribution to the production of direct photons, i.e. those photons which 
do not 
result from $\pi^0$ decay, etc. The spectrum of such photons 
provides, in principle, an excellent test of perturbative quantum 
chromodynamics
(QCD) because in this case one has a large number of data points against which 
theoretical predictions can be compared. This is in contrast to the prediction
of a decay rate, which is a single number. However it is well 
known{\cite{Schuler}}
that the photon spectrum $^3 S_1 \rightarrow \gamma \,+\, X$, calculated at 
leading order\cite{Brodsky}, is too hard when compared against experiment, 
both in $J/\Psi$ and 
$\Upsilon$ decays. Such calculations yield an almost linearly rising spectrum
in $z=2 E_{\gamma} / M$ with a sudden decrease at $z=1$. A next-to-leading
order calculation by Photiadis\cite{Photiadis} sums leading logs of the 
type $\ln(1-z)$ and
yields some softening. However, the peak is still too sharp and close to 
$z=1$. An earlier calculation by Field\cite{Field} predicts a much softer 
spectrum which fits the relatively recent data\cite{Bizzeti} quite well. 
This uses a
parton-shower Monte Carlo approximation wherein the two gluons recoiling 
against the direct photon acquire a non-zero invariant mass by radiating 
further bremsstrahlung gluons. This does not, therefore, qualify it as an
ab-initio perturbative QCD calculation. We note that in refs[2-4] the 
non-perturbative dynamics of the decaying hadron is described by a single 
parameter $\phi(0)$, the quark wavefunction at the origin. This leads to the 
assertion that the ratio of widths for decay into prompt photons and $l^+l^-$ 
pairs is independent of quark dynamics.

In this paper we compute the rate for $^3 S_1 \rightarrow \gamma \,+\, X$ 
taking into account the bound state structure of the decaying quarkonium 
state. We note that the description of hadron dynamics in this decay process 
by just $\phi(0)$ is correct {\it only} if one assumes that $Q$ and 
$\bar{Q}$ are 
exactly on-shell and at rest relative to each other. This assumption is
only approximately true -- heavy quarkonia are weakly bound  $Q \bar{Q}$ 
composites and $v^2/c^2$ is a small parameter. 
Improvement requires introduction of additional hadronic quantities,
which we identify within  the context of a systematically improvable 
gauge-invariant theory for quarkonium decays. This formalism has been 
recently applied to one and two particle decays\cite{Khan,Bodwin}. Here 
we apply
the method of ref\cite{Khan} to the more complicated three particle case 
and obtain the photon spectrum for the process 
$\Upsilon \rightarrow \gamma \,+\, 2g$. We find that 
inclusion of binding and relativistic effects via the two additional 
parameters,
$\epsilon_B/M$ and $\nabla^2 \phi(0)/M^2\phi(0)$, makes the computed spectrum 
softer for large $z$, ($z<0.9$). For still larger z, $0.9<z<1$, 
there are non-perturbative effects due to final-state gluon interactions which
cannot be reliably computed and which, therefore, we shall not address.   

\section*{FORMALISM}
The starting point of our approach is that the decay amplitude for 
$^3 S_1 \rightarrow \gamma + X$ is given by the sum of all distinct 
Feynman diagrams leading from
the initial to the final state. The first step is to write a 
given diagram in the form of a (multiple) loop integral. Consider, for example,
one of the six leading order diagrams (Fig. 1a). 
Omitting colour matrices and coupling constants for brevity, its contribution 
can be expressed as
\be
T^{\mu_1 \mu_2 \mu_3}_{o(1b)} = \int \frac{d^4k}{(2 \pi)^4} Tr \left[
\gamma^{\mu_2} S_F(k+s_2) \gamma^{\mu_1} S_F(k-s_3) \gamma^{\mu_3} M(k)\right].
\ee
$M(k)$ is the usual, but obviously non-gauge invariant zero-gluon, 
Bethe-Salpeter amplitude,
\be
M(k) = \int d^4\! x \,e^{i k\cdot x} \langle 0 |T [\bar{\psi} (-x/2) 
\psi(x/2) ] | P,\epsilon \rangle.
\ee
In equations 1-2, $x^{\mu}$ is the relative distance between quarks, 
$k^{\mu}$ is the relative momentum, $P^2=M^2$, and $s_i = q_i - \frac{1}{2}P$.
We define the binding energy as $\epsilon_B = 2 m - M$.
Provided all propagators are far off-shell, they may be expanded in the two 
small quantities $\epsilon_B/M$ and $k/M$.
This yields the expression,
\bea
T_o^{\mu_1 \mu_2 \mu_3} = &Tr& [  \langle 0 |\bar{\psi}\psi | P,\epsilon
\rangle h^{\mu_1 \mu_2 \mu_3}
+ \langle 0 |\bar{\psi} \twowaypartial_{\alpha}\psi | P,\epsilon 
\rangle \partial^{\alpha} h^{\mu_1 \mu_2 \mu_3}  \nonumber \\
&+& \langle 0 | \bar{\psi} \twowaypartial_{\alpha} \twowaypartial_{\beta}
\psi | P,\epsilon\rangle \frac{1}{2} \partial^{\alpha} \partial^{\beta} 
h^{\mu_1 \mu_2 \mu_3} + \dots ].
\eea
We have defined $\stackrel{\leftrightarrow}{\partial}^{\alpha} = 
\frac{1}{2} (\stackrel{\rightarrow}{\partial}^{\alpha}\!\!\!-\!
\stackrel{\leftarrow}{\partial}^{\alpha})$, and $h^{\mu_1 \mu_2 \mu_3}$ 
is the ``hard part" which combines terms from all six leading 
diagrams\footnote{
In actual fact, only three of the diagrams need to be evaluated 
because of time-reversal symmetry. This simplification halves the number
of diagrams in the one-gluon and two-gluon cases as well.}.
One can readily see that it is the sum of terms of
the type in eq.1 corresponding to different 
permutations of indices and momenta.
There are 12 one-gluon diagrams one of which is illustrated in fig 1b, which
must be added as corrections to the no-gluon amplitude. These all have the 
general form
\be
T_1^{\mu_1 \mu_2 \mu_3} = \int \frac{d^4 k}{(2 \pi)^4} 
\frac{d^4 k'}{(2 \pi)^4} Tr \,\, M^{\rho}(k,k') H_{\rho}^{\mu_1 \mu_2 \mu_3}
(k,k'),
\ee
where $M_{\rho}(k,k')$ is a generalized B-S amplitude,
\be
M^{\rho}(k,k') = \int d^4 \!x \, d^4 \! z \,e^{i k\cdot x} e^{i k'\cdot z} 
\langle 0 |T [\bar{\psi} (-x/2) A^{\rho}(z) \psi(x/2) ] | P,\epsilon 
\rangle.
\ee
The gluon which originates from the blob is part of the $Q \bar{Q} g$
Fock-state component of the meson and has its momentum $k'$ bounded by $R^{-1}
\simle k' \ll M$, where R is the meson's spatial size. Hence it is to be 
considered soft on 
the scale of quark mass. Again, one may expand the propagators in $H_{\rho}
^{\mu_1 \mu_2 \mu_3}(k,k')$  about $k=k'=0$ to get
\be
T_1^{\mu_1 \mu_2 \mu_3} = Tr[ M^{\rho} H_{\rho}^{\mu_1 \mu_2 \mu_3} 
+ M^{\rho,\alpha} \partial_{\alpha} H_{\rho}^{\mu_1 \mu_2 \mu_3} 
+ M'^{\rho,\alpha} \partial'_{\alpha} H_{\rho}^{\mu_1 \mu_2 \mu_3} + \dots ],
\ee
where,
\bea
M^{\rho} &=& \langle 0 |\bar{\psi} \psi A^{\rho}| P,\epsilon 
\rangle \nonumber \\M^{\rho,\alpha} &=& \langle 0 |\bar{\psi} 
\twowaypartial^{\alpha}\!\!\!\psi A^{\rho} 
|P,\epsilon \rangle \nonumber \\
M^{' \rho,\alpha} &=& \langle 0 |\bar{\psi}  
\psi i\!\!\stackrel{\rightarrow}{\partial}^{\alpha}\!\!\!\!\!A^{\rho}| P,
\epsilon \rangle ,
\eea
The derivative $\twowaypartial^{\alpha}$ acts only upon the quark operators.

The two soft-gluon contribution (figs.1c-1d) to the amplitude is handled 
similarly but is more complicated. The matrix
elements which enter into $T_2^{\mu_1 \mu_2 \mu_3}(k)$  are of the type
$<0|\bar{\psi}\psi A^{\rho} A^{\lambda}|P,\epsilon>$ and so on; we 
shall not list these explicitly here since the principle
is rather clear. It is straightforward to show that in the sum 
$T_0+T_1+T_2$, all $\partial's$  combine with $A's$  to yield covariant 
derivatives and/or field strength tensors,
\bea
(T_0+T_1+T_2)^{\mu_1 \mu_2 \mu_3} = Tr [ \langle 0 |\bar{\psi}\psi | P,\epsilon
\rangle h^{\mu_1 \mu_2 \mu_3} 
+ \langle 0 |\bar{\psi} i\!\!\stackrel{\leftrightarrow}{D}_{\alpha}\!\psi 
|P,\epsilon\rangle \partial^{\alpha} h^{\mu_1 \mu_2 \mu_3}  \nonumber \\
+ \langle 0 |\bar{\psi} i\!\!\stackrel{\leftrightarrow}{D}_{\alpha}\! 
i\!\!\stackrel{\leftrightarrow}{D}_{\beta}\!
\psi | P,\epsilon\rangle \frac{1}{2} \partial^{\alpha} \partial^{\beta} 
h^{\mu_1 \mu_2 \mu_3}  
+\langle 0 |\bar{\psi} F^{\alpha \beta}\psi | P,\epsilon \rangle 
\frac{i}{2} \partial'_{\alpha}H_{\beta}^{\mu_1 \mu_2 \mu_3}+ \dots ].
\eea
The above is a sum of terms, each of which is the product of a soft hadronic
matrix element and a hard perturbative part.

To proceed, one can perform a Lorentz and CPT invariant decomposition of each
of the hadronic matrix elements in Eq 10. 
This is somewhat complicated\cite{Yusuf} and involves a large number of 
constants which
characterize the hadron. Considerable simplification results from choosing the
Coulomb gauge, together with the counting rules of Lepage et. al
\cite{Lepage}. The upshot 
of using this analysis is that, in this particular gauge, the gluons 
contribute
at $O(v^3)$ to the reaction $^3 S_1 \rightarrow \gamma + X$, and hence can
be ignored. Even this leaves us with too many parameters, and forces us to
search for a dynamical theory for the $1^{--}$ quarkonium state. We shall 
assume, in common with
many other authors, that the Bethe-Salpeter equation with an instantaneous 
kernel does provide an adequate description. This has been conveniently 
reviewed by Keung and Muzinich\cite{Keung} and we adopt their 
notation\footnote{We find 
the analysis of ref\cite{Keung} to be wanting because it 
does not properly deal with the issue of gauge-invariance of the meson state.
Further, while the binding energy is taken into account, the wavefunction 
corrections - which are essentially short-distance or relativistic 
effects - are not.}. The 
momentum space B-S amplitude $\chi (p)$ satisfies the homogeneous equation,
\begin{equation}
\chi (p) = i G_0(P,p) \int \frac{d^4 p'}{(2 \pi)^4} \;\;K(P,p,p') \;\;\chi(p'),
\end{equation}
which, after making the instantaneous approximation 
$K(P,p,p')=V(\vec{p},\vec{p'})$ and reduction to the non-relativistic limit 
yields,
\begin{equation}
\chi (p) = \frac{M^{1/2} (M-2E) (E+m-\vec{p}.\vec{\gamma} )\slashepsilon 
(1-\gamma_0) (E+m-\vec{p}.\vec{\gamma}) \phi (| \vec{p} |)}{4 E (E+m) 
( p^0+\frac{M}{2}-E+i\epsilon ) ( p^0-\frac{M}{2}+E-i\epsilon )}.
\end{equation}
The scalar wavefunction $\phi ( | \vec{p} |)$ is normalized to unity,
\begin{equation}
\int \frac{d^3p}{(2 \pi)^3} | \phi (| \vec{p} |) |^2 =1,
\end{equation}
and,
\begin{equation}
E = \sqrt{\vec{p}^2 +m^2}.
\end{equation}
Fourier transforming $\chi(p)$ to position space yields 
$\langle 0 | \bar{\psi}(-x/2) \psi (x/2) | P\rangle$ from which, by tracing 
with appropriate gamma matrices, the coefficients below can be
extracted. So finally, to $O(v^2)$, one has a rather simple result,
\bea
\label{me}
\langle 0 | \bar{\psi} \psi |P,\epsilon\rangle & = & \frac{1}{2} M^{1/2} 
\! \left( 1\!+\!\frac{\nabla^2}{M^2} \right) \phi \left( 1\!+\!\frac{\slashP}
{M} \right) 
\slashepsilon-\frac{1}{2} M^{1/2} \! \frac{\nabla^2 \phi}{3 M^2} \left( 
1\!-\!\frac{\slashP}{M} \right)  \slashepsilon, \nonumber \\
\langle 0 |\bar{\psi}  \twowaypartial_{\alpha}\! \psi | P,\epsilon \rangle  
&=& -\frac{1}{3}M^{3/2} \frac{\nabla^2\phi}{M^2} \epsilon^{\beta} 
\left[ -g_{\alpha\beta} + i \epsilon_{\mu\nu\alpha\beta} \frac{P^{\nu}}{M} 
\gamma^{\mu} \gamma^5\right], \nonumber \\
\langle 0 |\bar{\psi}  \twowaypartial_{\alpha}  \twowaypartial_{\beta}\!
\psi | P,\epsilon
\rangle &=& \frac{1}{6}M^{5/2} \frac{\nabla^2 \phi}{M^2} \left( g_{\alpha
\beta} -\frac{P_{\alpha}P_{\beta}}{M^2} \right)  \left( 1 \!+\! \frac{\slashP}
{M} \right) \slashepsilon.
\eea

\section*{DECAY RATE}
All the ingredients are now in place to calculate the decay 
$\Upsilon \rightarrow \gamma \,+\, 2g$. In squaring the amplitude obtained 
by substituting Eqs.13 into Eq.8, terms involving the product of 
$\epsilon_B$ and $\nabla^2 \phi$ may be neglected.
We assume the emitted gluons to be 
massless and transverse, and to decay with unit probability into hadrons.
Polarizations of the final-state particles are summed over, and the spin
states of the initial meson are averaged over. Summing over final-state 
colors yields $2/3$, and one must include a factor of $1/2$ for identical 
gluons. The Lorentz invariant phase-space factor for 3 massless particles 
has a standard expression\cite{Itz} which is best expressed in terms of the
dimensionless energy fractions $x_i=2E_i/M$ which satisfy  $x_1+x_2+x_3=2$.
The variables $s,t,u$ are symmetric functions of $x_i$,
\bea
s &=& (P-q_1)^2 \nonumber \\
&=& M^2(1-x_1),
\eea
Ignoring radiative radiative corrections for the
moment, a tedious calculation\footnote{
We used Mathematica\cite{Wolfram}, supplemented by the HIP package\cite{HIP}, 
for computation  of traces and simplification of algebra} yields,
\be   
\frac{d^2\Gamma}{dx_1\,dx_2}= \frac{256}{9} e_q^2\alpha_s^2 \alpha_e\frac
{|\phi(0)|^2}{M^2}
\left[ \eta_0 f_0(s,t,u)+\eta_B f_B(s,t,u)+\eta_W f_W(s,t,u) \right].
\ee
$e_q$ is the quark charge and,
\be
\eta_0=1, \,\, \eta_B=\frac{\epsilon_B}{M}, \,\, \eta_W=\frac
{\nabla^2\phi}{M^2\phi}.
\ee
The function $f_0$ provides the standard, leading order result:
\be
f_0(s,t,u) = \frac{M^4\left(s^2\, t^2+t^2\,u^2+u^2\, s^2 +M^2\,s\,t\,u\right)}
{(s-M^2)^2 (t-M^2)^2(u-M^2)^2}.
\ee
The binding energy and wavefunction corrections, $f_B$ and $f_W$ respectively,
are more complicated:
\bea
f_B(s,t,u) &=& \frac{M^4}{2D}\left[- 7\,s\,t\,u\,\left(s^4+t^4+u^4\right)+
7\,M^2\,\left(s^3 t^3+t^3 u^3 +u^3 s^3\right)\right.
\nonumber \\ &+&\left(s^2 t^2+ t^2 u^2 + u^2 s^2\right)\left
(s^3+t^3+u^3+15\,s\,t\,u\right)
\nonumber \\ &+& \left.M^2\,s\,t\,u\left(s^3+t^3+u^3\right)+29 M^2 s^2 t^2 u^2
 \right], \nonumber \\
\nonumber \\
f_W(s,t,u) &=& \frac{M^4}{3 D}\left[141\,s\,t\,u\,\left(s^4+t^4+u^4\right)-
85\,M^2\,\left(s^3 t^3+t^3 u^3 +u^3 s^3\right) \right.
\nonumber \\ &-&27\,\left(s^2 t^2+ t^2 u^2 + u^2 s^2\right)\left(s^3+t^3+u^3+
\frac{205}{27}\,s\,t\,u\right)
\nonumber \\ &-&\left. 139\,M^2\,s\,t\,u\,\left(s^3+t^3+u^3\right)-463\, 
M^2\, s^2 t^2 u^2\right].
\eea
The denominator $D$ is,
\be
D = (s-M^2)^3 (t-M^2)^3 (u-M^2)^3.
\ee
Integrating over the energies of the outgoing gluons for a fixed photon energy
yields
\be
\frac{d\Gamma}{dz} = \frac{256}{9} e_q^2 \alpha_e \alpha_s^2 
\frac{|\phi(0)|^2}{M^2} \left[\eta_0 F_0(z)+\eta_B F_B(z)+
\eta_W F_W(z)\right],
\ee
where $z=2E_{\gamma}/M$ and,
\bea
F_0 &=& [1+4\xi-2\xi^3-\xi^4-2\xi^5+2\xi(1+2\xi+5\xi^2)\log \xi]/(1-\xi)^2 
(1+\xi)^3,\nonumber \\ 
\nonumber \\
F_B &=& [2-16\xi+10\xi^2-48\xi^3 -10\xi^4+64\xi^5-2\xi^6 \nonumber \\
&+& (1-3\xi+14\xi^2-106\xi^3+17\xi^4 -51\xi^5)\log \xi]/2\,(1-\xi)^3 
(1+\xi)^4, \nonumber \\
\nonumber \\
F_W &=&[-26+14\xi-210\xi^2+134\xi^3+274\xi^4-150\xi^5-38\xi^6+2\
xi^7 \nonumber \\
&-& (27+50\xi+257\xi^2-292\xi^3+205\xi^4-78\xi^5-41\xi^6)\log \xi]/\nonumber \\
&&3(1-\xi)^3 (1+\xi)^5.
\eea
In the above, $\xi=1-z$. The integrated decay width is\footnote
{Note that Eq.22 does not take into account non-perturbative effects\
cite{Voloshin} which are significant in the
part of the phase space where one of the quark propagators become soft, and 
where the gluon vacuum condensate plays a role.},
\be
\Gamma_{1^{--} \rightarrow \gamma + 2 \, g} = \frac{128}{9}(\pi^2-9) e_q^2 
\alpha_e \alpha_s^2 \frac{|\phi(0)|^2}{M^2}\left(1+a \frac{\alpha_s}{\pi} -
1.03 \eta_B + 19.34 \eta_W\right).
\ee
Where we have included the radiative corrections of $O(\alpha_s)$ which are 
of the same order in $v^2/c^2$ as the other corrections, and were 
calculated\cite{Mackenzie} many years ago,
\be
a = \beta_0 \ln(\mu/m_Q) - 4.37 - 0.77 n_f,
\ee
where $\beta_0 = 11 -2n_f/3$. The parameters $\eta_W$ and $\eta_B$ are 
independent of each other in the 
present treatment. We note, however, that if we impose the condition 
$\eta_W = \half \eta_B$ then the result Eq. 3.5 of Keung and Muzinich[10] is 
precisely recovered. This latter condition is equivalent to $\frac{1}{M}
\nabla^2\phi(0)= \half \epsilon_B \phi(0)$, which is the Schr\"{o}dinger 
equation for quark relative motion in a potential which vanishes at 
zero relative separation. It is also worthy of note that the same condition 
emerges as a 
renormalization condition in the treatment of positronium by Labelle et 
al\cite{Labelle}
(see their equations 11 and 12). However in our treatment there is no principle
which apriori constrains $\eta_B$ to bear a fixed relation to $\eta_W$ and
therefore both will be considered adjustable parameters.

The application of Eq.22 must be done with caution because extraction of the 
direct photon decay rate from the data requires an extrapolation down to 
small photon energies. But in this energy range the prompt photons are heavily 
contaminated by photons from $\pi^0$ decays. 
A numerical estimate of the correction factors 
requires the value of $\eta_B$ and $\eta_W$. We have chosen  $m_b=4.5$
which gives $\eta_B=-0.048$. If we take $\alpha_S=0.20$ then $\eta_W$ can be 
fixed by using the experimentally known numbers\cite{Review},
\bea
\Gamma(\Upsilon \rightarrow 2g+\gamma) &=& 1.28 \pm 0.10 \:\: KeV.\nonumber \\
\Gamma(\Upsilon \rightarrow l\,\bar{l}) &=& 1.34 \pm 0.04 \:\: KeV.
\eea
This gives a range of values for $\eta_W$. We have plotted the graphs in
fig.2 at $\eta_W=-0.0059$.
The binding, $F_B(z)$, and wave-function, $F_W(z)$,
correction terms tend to cancel each other over part of the z region.  
The effect of final-state interaction 
corrections can be reasonably well estimated[3] provided one stays
away from the end-point $z=1$. In fig.2 we compare the data, 
taken from Ref[5], with the prediction of our model appropriately folded with 
the experimental photon energy resolution (assumed to be Gaussian). 
The effect of the binding 
and wavefunction corrections calculated in this work is sizeable, and tends to
to increase the photon rate in the middle z range and to lower it for larger 
z. While this appears to be in the right direction, it would be highly 
desirable to have more precise data.

\section*{SUMMARY}

The approach taken in this paper for calculating 
the amplitude for  $\Upsilon \rightarrow \gamma \,+\, X$ 
is to take the sum of all distinct Feynman diagrams   
leading from the initial quarkonium state to the final state. Each diagram 
is put into the form of a (multiple) loop integral with a kernel which is a 
product of a hard part and a soft part. The hard part is treated with 
perturbative QCD, and the soft part is analyzed into its different components
with the use of Lorentz, $\cal C$, and $\cal P$ symmetries. Use of the QCD 
equations of 
motion enables separation of these components according to their importance 
in powers of $v$. At the last step, a specific commitment to dynamics is made 
and the B-S equation is used to express the components in the form of 
wavefunctions. However, the un-regularized value of  $\nabla^2 \phi(0)$ is 
singular at the origin $\nabla^2 \phi(0) \sim  M \phi(0)/r$. As is clear from 
the uncertainty principle, the local kinetic energy becomes very large at 
short distances and the expansion in powers of $v$ breaks down. This 
difficulty 
was circumvented by imagining that annihilation takes place in a diffused 
region of size $O(1/m)$, i.e., that $\phi(0)$ and $\nabla^2 \phi(0)$ are 
quantities renormalized
at this scale, and to be considered as adjustable parameters. The numerical
investigation we undertook showed that varying these within reasonable 
limits led to substantial improvement in the intermediate z region but
was insufficient to reproduce the data near $z=1$, once again
underscoring the importance of final-state interactions between collinear 
gluons.

\newpage
 \centerline{\bf Acknowledgements}
This work was supported in parts by funds provided by NSF Grant No. 
INT-9122027. M.A.Y gratefully acknowledges the financial support from the 
Pakistan Science Foundation for his doctoral research .

\newpage 

\begin{center}
{\Large Figure Captions}
\end{center}
\begin{tabbing}
1. \= a) \= One of the six leading order diagrams.\= \\
   \> b) \> One of the 12 one-gluon diagrams.     \> \\
   \> c) \> One of the 24 two-gluon diagrams .    \> \\
   \> d) \> One of the 12 gluon self-coupling diagrams.
\end{tabbing}

{\raggedright 2. The photon spectrum as a function of $z$, folded with the 
experimental photon energy resolution. The dotted line is 
the zeroth order QCD result, the dashed line incorporates the binding and 
wavefunction corrections, with $\eta_B=-.048$ and $\eta_W=+0.0059$ .
The solid line is the final result including final-state interaction of 
Ref[3].}

\newpage
\begin{figure}
\begin{center}
\psfig{file=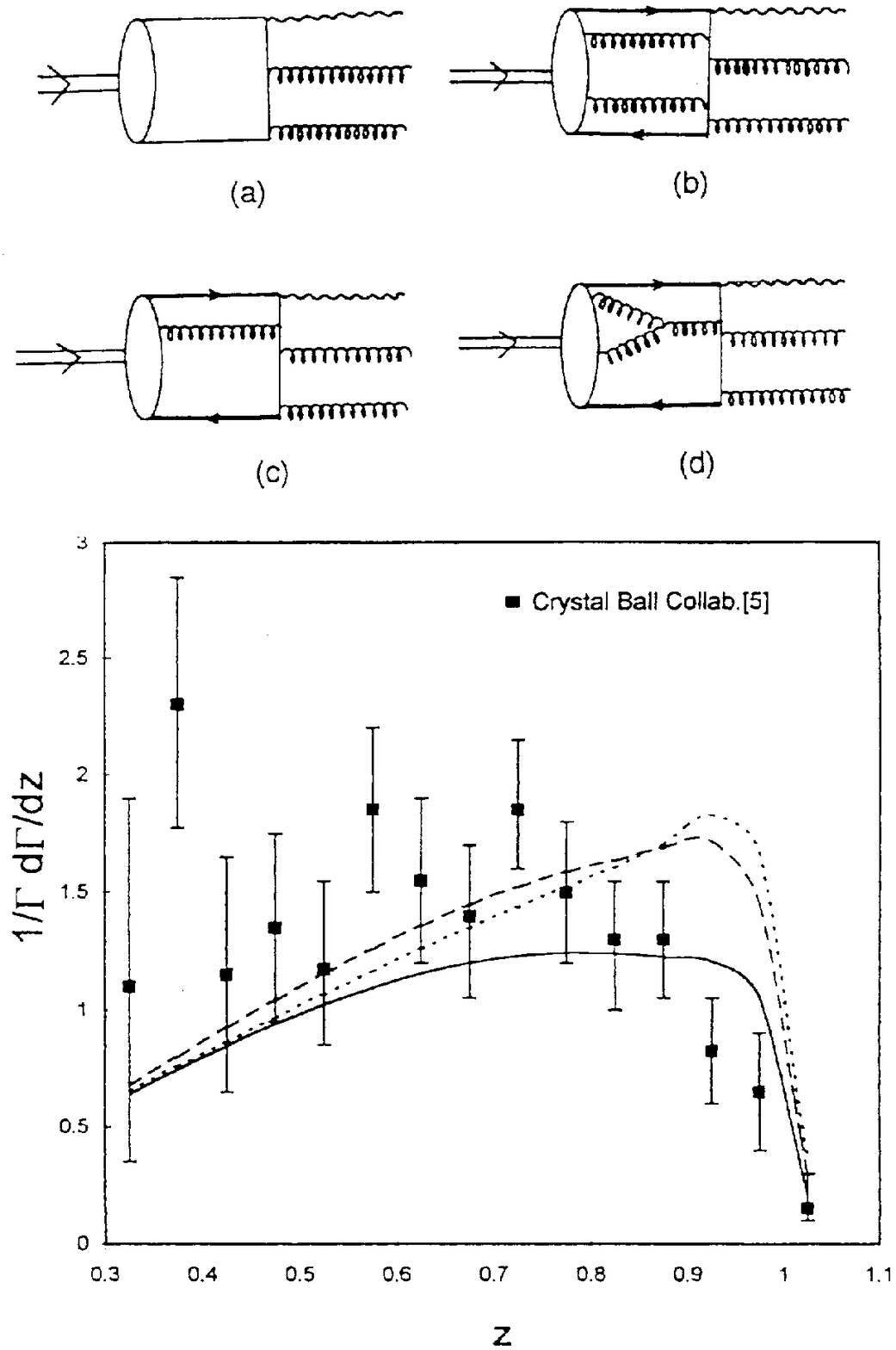,width=\textwidth}
\end{center}
\end{figure}


\begin{thebibliography}{99}
\bibitem{Schuler} For a review, see Gerhard A. Schuler, ``Quarkonium 
Production And Decays",  CERN-TH. 7170/94, submitted to Phys. Rep.
\bibitem{Brodsky} S.Brodsky, G.P.Lepage, and P.B.Mackenzie, Phys.Rev. 
{\bf D28} 228 (1983)
\bibitem{Photiadis} Douglas M. Photiadis, Phys.Lett. {\bf B164} 160 (1985).
\bibitem{Field} R.D.Field, Phys.Lett. {\bf B133} 248 (1983).
\bibitem{Bizzeti} Bizzeti et al, Crystal Ball Collaboration, Phys.Lett. 
{\bf B267} 286 (1991).
\bibitem{Khan} Hafsa Khan and Pervez Hoodbhoy,  Phys.Rev. {\bf D53} 
2564 (1996).
\bibitem{Bodwin} G.T. Bodwin, E.Braaten, and G.P.Lepage, Phys.Rev. {\bf D51} 
1125 (1995).
\bibitem{Yusuf} M.A.Yusuf, thesis in preparation, Quaid-i-Azam University, 
Islamabad, Pakistan.
\bibitem{Lepage} G.P.Lepage, L.Magnea, C.Nakleh, U.Magnea, K.Hornbostel, 
preprint CLNS 92/1136, OHSTPY-HEP-T-92-001, Feb. 1992.
\bibitem{Keung} W.Y.Keung and I.J.Muzinich, Phys.Rev {\bf D27} 1518 (1983).
\bibitem{Mackenzie} P.B.Mackenzie and G.P.Lepage, Phys.Rev.Lett {\bf 47} 1244 (1981),\\ 
W.Kwong, P.B.Mackenzie, R.Rosenfield, and J.L.Rosner, Phys.Rev. {\bf D37} 3210 (1988).
\bibitem{Wolfram} S.Wolfram, ``MATHEMATICA - A System for Doing Mathematics by Computer", Second edition, Addison-Wesley Publishing Company, 1991.
\bibitem{Itz} C.Ityzkson and J.Zuber, ``Quantum Field Theory", p.237, 
McGraw-Hill, 1980.
\bibitem{HIP} A. Hsieh and E. Yehudai, ``HIP: Symbolic high-energy physics 
calculations", preprint SLAC-PUB 5576.
\bibitem{Labelle} P. Labelle, G.P. Lepage and U. Magnea, 
Phys.Rev.Lett. \underline{72}, 2006 (1994)
\bibitem{Voloshin} M.B. Voloshin, Sov.J.Nucl.Phys. {\bf40} 662 (1984).
\bibitem{Review} Review of Particle Properties, Phys.Rev. {\bf D50} 1177 
(1994).
\end{thebibliography}
\end{document}